# Queues, Stacks, and Transcendentality at the Transition to Chaos

Cristopher Moore[1] and Porus Lakdawala[2]

[1] Santa Fe Institute, 1399 Hyde Park Road, Santa Fe NM 87501
`moore@santafe.edu`
[2] Homi Bhabha Center for Science Education, V.N. Purav Marg, Mumbai 400088, India
`porus@theory.tifr.res.in`

**Abstract.** We examine the one-humped map at the period-doubling transition to chaos, and ask whether its long-term memory is stack-like (last-in, first-out) or queue-like (first-in, first-out). We show that it can be recognized by a real-time automaton with one queue, or two stacks, and give several new grammatical characterizations of it. We argue that its memory has a queue-like character, since a single stack does not suffice. We also show that its dynamical zeta function, generating function and growth function are transcendental. The same results hold for any period-multiplying cascade. We suggest that transcendentality might be a sign of dynamical phase transitions in other systems as well.

## 1 Introduction

One of the basic distinctions in physics and dynamical systems is between finite-state processes, characterized by short-range, exponentially decaying correlations, and processes with an infinite amount of memory, causing long-range correlations such as power laws.

But we can draw a finer distinction based on what kind of long-term memory a system has. Two of the basic data structures studied in computer science are the *stack*, in which the most recent symbols pushed on must be read and removed before older symbols can be read (last-in, first-out), and the *queue*, in which symbols are read in the same order they are entered (first-in, first-out). These are two fundamentally different ways for a system to depend on its history.

For instance, a stack can recognize the set of palindromes, such as *abccba*, while a queue prefers to repeat things in the same order, as in *abcabc*. Both have long-range correlations; the difference is that they are nested in the first case and cross each other in the second, as shown in figure 1. While a stack pushes and pops at the same end, a queue pushes symbols on one end (the right, say) and pops them off the other.

In this paper, we examine the symbolic dynamics of the one-humped map at the period-doubling transition to chaos. We show that it can be recognized in real time by an automaton with access to two stacks, or one queue. We review and extend previous descriptions of it as an indexed context-free language, and also show that it can be generated by a *breadth-first context-free grammar*. Since it is more easily recognized with queues than with stacks, we argue that the system's memory has a queue-like character.

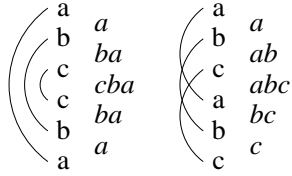

**Fig. 1.** Both stacks and queues have long-term memory — but stacks have nested correlations, while queues have crossing ones.

In addition, we show that various functions associated with the system, including its dynamical zeta function and the generating and growth functions of its symbolic dynamics, are transcendental. We show that this is true for period-multiplying cascades and general, and speculate as to whether these characteristics might be common to other kinds of dynamical phase transitions.

A preliminary version of this work appeared in [14].

## 2 The period-doubling cascade

Consider the family of functions on the unit interval $[0, 1]$

$$F_\mu(x) = 4\mu x(1 - x)$$

As $\mu$ increases, $F_\mu$ undergoes a series of bifurcations forming stable periodic orbits of period 1, 2, 4, ... Each new orbit is formed by a period-doubling transition from the one before it, which then becomes unstable.

At any given stage in this process, there is a stable periodic orbit of period $2^n$ which attracts almost all initial points, and unstable orbits with periods $2^k$ for all $k < n$. All these bifurcations accumulate at the transition to chaos, at which all these orbits are unstable and an aperiodic attractor (of period $2^\infty$) appears.

If we label the two halves of the unit interval 0 and 1 for $x \leq 1/2$ and $x > 1/2$ respectively, any point $x$ can be assigned a sequence or 'itinerary' in the following way:

$$a : I \to \{0, 1\}^{\mathbb{N}-\{0\}} : a_t(x) = \begin{cases} 0 \text{ if } F_\mu^t(x) \leq 1/2 \\ 1 \text{ if } F_\mu^t(x) > 1/2 \end{cases}$$

Then the *language* or set of sequences

$$L = \{a(x) \mid x \in [0, 1]\}$$

represents the *symbolic dynamics* [10] of $F_\mu$.

As we go through this cascade of bifurcations, the symbolic dynamics change accordingly. For $\mu < 1/2$, all points have images in the left half of the interval:

therefore the only possible itinerary is $000\cdots = 0^*$, where the $*$ operator indicates 0 or more repetitions. As $\mu$ increases, we get an unstable fixed point at 0 and a stable one which migrates into the right half; the symbolic dynamics is now $0^*1^*$, indicating that once we cross over into the right half-interval, we can never come back.

At the next bifurcation, the stable fixed point splits into a period-2 orbit and becomes unstable. As $\mu$ increases further, one of these two points crosses back onto the left side, giving the orbit an itinerary of $(01)^*$; the symbolic dynamics now becomes $0^*1^*(01)^*$. As this process continues, we get a sequence of languages:

$$L_0 = 0^* w_0^*$$
$$L_1 = 0^* w_0^* w_1^*$$
$$L_2 = 0^* w_0^* w_1^* w_2^*$$
$$\vdots$$

where $L_n$ is the language when the period-$2^n$ orbit is stable. The itineraries of the periodic orbits of period $2^n$ are $w_0 = 1$, $w_1 = 10$, $w_2 = 1011$, $w_3 = 1011\,1010$, and so on. Note that $w_{n+1}$ consists of two copies of $w_n$ with the last symbol complemented; equivalently, $w_{n+1} = R(w_n)$ where $R$ is the renormalization

$$R : 0 \to 11,\ 1 \to 10$$

As we approach the transition to chaos, the stable periodic points approach an attractor whose itinerary is the fixed point of $R$, the *Morse sequence* $w_\infty = 1011\,1010\,1011\,1011\cdots$. The symbolic dynamics is then

$$L_\infty = 0^* w_0^* w_1^* w_2^* w_3^* \cdots = \cup_{n=0}^\infty L_n$$

Note that unlike some other studies of this system [6, 12], we are including the transient part of the dynamics, rather than just the set $\{w_n\}$ of periodic orbits or the attractor $w_\infty$.

Each $L_n$ is a *regular* language [11] in that it is generated by a finite-state Markov process, or recognized by a finite-state machine. However, $L_\infty$ is not regular, since it has an infinite number of inequivalent states. This was first pointed out by Grassberger [9] who gave an infinite-state transition graph for $L_\infty$ similar to that shown in figure 2. This means that the system has an infinite amount of memory. But what kind?

## 3 Languages and automata for $L_\infty$

### 3.1 One stack can't do it

Our approach to automata and grammars will be somewhat informal. We recommend [11] as an introduction.

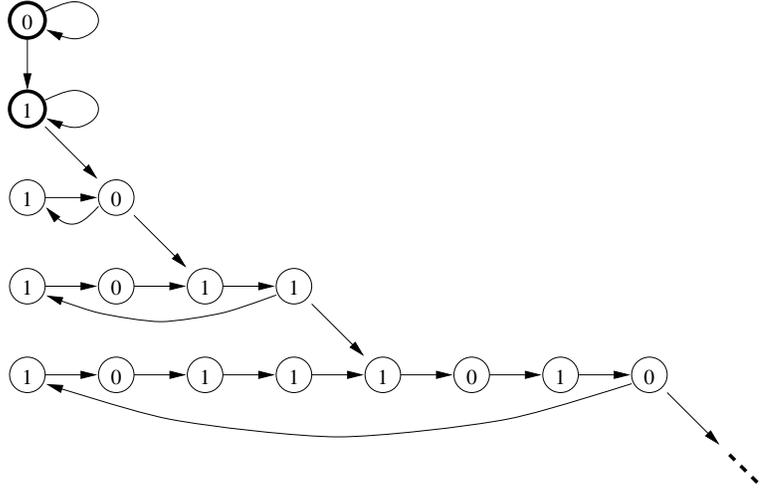

**Fig. 2.** The infinite-state transition diagram for $L_\infty$. We can start from either bold node.

A *push-down automaton* (PDA) is a machine with a single stack and a finite-state control. PDAs recognize the *context-free* languages (CFLs), so-called because they can be generated by a grammar in which symbols can be transformed into strings regardless of the adjacent symbols. The canonical example is the Dyck language $L = \{\epsilon, (), (()), ()(), \ldots\}$ of properly nested words of parentheses (here $\epsilon$ is the empty word). A PDA can recognize $L$ by pushing a symbol on the stack when it sees a '(', popping when it sees a ')', starting and ending with an empty stack, and refusing to pop an empty stack. The corresponding grammar

$$I \to (I)I, \epsilon$$

generates $L$ from the initial symbol $I$, where the comma indicates multiple options for the production rule. Here $I$ is a *variable* used during the derivation process, but all final words must consist only of the *terminals* '(' and ')'.

This grammar is *unambiguous*, in that each word in $L$ has a unique derivation. Since $L$ has a grammar of this kind, it is called an unambiguous CFL.

One useful property of context-free languages is the Pumping Lemma [11]: any sufficiently long word in $L$ can be written $\alpha\beta\gamma\delta\mu$ such that $\alpha\beta^k\gamma\delta^k\mu$ is also in $L$ for all $k \geq 0$. For the Dyck language, $\beta$ and $\delta$ are simply the left and right parentheses.

Crutchfield and Young [6] point out that the Pumping Lemma can be used to show that several languages related to $L_\infty$, including $\{w_n\}$ and the set of substrings of $w_\infty$, are not CFLs. On the other hand, $L_\infty$ obeys the Pumping Lemma since initial substrings of a word can be repeated and considered part of the transient. In order to show that $L_\infty$ is not a CFL, we need to use Ogden's Lemma [15], in which we can mark particular symbols in a word and demand that they be among those pumped. Since the periodicity of a word in $L_\infty$ must increase from left to right, symbols near the end of $w_n$ for large $n$ can't be repeated, and $L_\infty$ is not a CFL. The actual proof of this is given in the Appendix.

A more powerful kind of machine is a *stack automaton* [11], which can delve into the stack in a read-only mode. A *one-way* stack automaton (1NSA) is one which can only read its input from left to right. Lakdawala [12] pointed out that the set $\{w_n\}$ cannot be recognized by a 1NSA, since another pumping lemma of Ogden [16] shows that any such language cannot be more than quadratically sparse, i.e. there must be sequences of words $w_n$ of length $|w_n| < bn^2$ for some $b$. Since the set of periodic orbits $\{w_n\}$ is exponentially sparse, it cannot be recognized by a 1NSA. Extending this proof to $L_\infty$ seems fairly difficult.

### 3.2 A stack of stacks can

An *indexed context-free* language [1] is a context-free language using indexed symbols, which have attached *indices* which are strings in another alphabet. When a symbol produces a string of symbols through a context-free production rule, the indices are copied to all the 'daughter' symbols. In addition, symbols can be pushed or popped from the index, which acts like a stack. The final product must consist of symbols with empty indices.

Indexed CFLs are recognized by *nested stack automata* [2]. These are considerably more powerful than PDAs; in addition to reading the top symbol, they can move down into the stack in a read-only mode. They can also create a new sub-stack within a stack, and so on to any number of levels, but cannot move up past a sub-stack until it is empty. In essence, they have a stack of stacks.

We need concern ourselves here just with indexed languages where the index alphabet consists of one symbol, say $x$. Then we can think of the indices as non-negative integers, by using the notation $A_i$ for $A$ indexed by the string $x^i$.

Consider the following indexed context-free grammar, in which $1_n$ produces $w_n$, the itinerary of the period-$2^n$ orbit:

$$1_n \to 1_{n-1} 0_{n-1}$$
$$0_n \to 1_{n-1} 1_{n-1}$$

Since the set $\{w_n\}$ of periodic orbits is formed by iterating the renormalization $R$, we can also think of it as a *0L-system* [11]. Indexed grammars of this kind for $\{w_n\}$ were given in [6] and [12].

To produce blocks of $w_n$'s with increasing $n$ we add the productions

$$I \to 0I, B_0$$
$$B_n \to 1_n, 1_n B_n, B_{n+1}$$

This grammar produces $L_\infty$. But to describe the symbolic dynamics completely, we really need to generate $\text{INIT}(L_\infty)$, the set of all initial substrings of $L_\infty$, that don't necessarily end in a complete copy of some $w_n$.

Since the indexed context-free languages form an *full abstract family of languages*, $\text{INIT}(L_\infty)$ is automatically also indexed context-free [11]. By adding a

few new symbols and productions, we can give the following grammar for it explicitly:

$$I \to \epsilon,\ 0I,\ B_0$$
$$B_n \to 1_n B_n,\ B_{n+1},\ 1'_n$$
$$1_n \to 1_{n-1} 0_{n-1}$$
$$0_n \to 1_{n-1} 1_{n-1}$$
$$1'_n \to 1_{n-1} 0'_{n-1},\ 1'_{n-1}$$
$$0'_n \to 1_{n-1} 1'_{n-1},\ 1'_{n-1}$$

Here the last block symbol $1'_n$, of which each derivation tree has one running down its right edge, generates either $w_n$ or an incomplete initial substring of it. It is easy to show that this grammar is unambiguous.

### 3.3 One queue can too

Cherubini et al. [4] define a *real-time queue automaton* as a finite-state automaton with access to a finite number of queues, in which it is allowed to make exactly one computation step per input symbol. The educated reader may know that, given unlimited time, a machine with a queue can simulate a universal Turing machine with quadratic slowdown. However, restricting it to real time creates an interesting class of languages.

In this section, we will show that $L_\infty$ is in the class $\mathbf{QA}_1$, recognizable by a real-time deterministic queue automaton with one queue. The idea is simply to store the current periodic orbit $w_n$ on the queue, and apply the renormalization when necessary to expand it.

Our automaton will have five states: `zero`, `repeat`, `expand`, `expand2`, and `reject`. Our queue alphabet will have two unmarked and two marked symbols, $\{0, 1, 0', 1'\}$. We will used the marked symbols $0'$ and $1'$ to mark when the queue comes to the end of some $w_n$. A word will be accepted as long as the automaton does not enter the `reject` state during the input process.

In our notation, we will use $a$ and $b$ as input symbols and $q$ as the leftmost symbol on the queue. We will use $R$ for a version of the renormalization which carries marks to the second symbol: $0 \to 11$, $1 \to 10$, $0' \to 11'$, $1' \to 10'$. We will ignore marks when testing equations such as $a = q$ or $ab = R(q)$. Then the automaton's program is given in table 1.

We start out in the `zero` state with an empty queue, and read 0s until we reach the first 1, whereupon we initialize the queue with $w_0 = 1'$.

The `repeat` state simply matches the input symbol to the leftmost symbol on the queue, pops it, and pushes it on the queue's right end, so that we cycle through the word $w_n$ stored in the queue. Since $w_{n+1}$ consists of two copies of $w_n$ with the last symbol complemented, `repeat` enters the `expand` state if the last (marked) symbol of $w_n$ does not match the input. If there is a mismatch before the end of $w_n$, there has been a mistake, and the machine rejects.

```
zero:
    – If a = 0, then stay in zero
    – If a = 1, then push 1' on the queue and go to repeat

repeat:
    – If a = q, then pop q, push q, and stay in repeat
    – If a ≠ q and q is marked, then pop q, push q, and go to expand
    – If a ≠ q and q is unmarked, then reject

expand:
    – If a is the first symbol of R(q), then go to expand2
    – If a is not the first symbol of R(q), then reject

expand2:
    – If a is the second symbol of R(q) and q is unmarked, then pop q, push
      R(q) and stay in expand
    – If a is the second symbol of R(q) and q is marked, then pop q, push R(q),
      and go to repeat
    – If a is not the second symbol of R(q) and q is marked, then pop q, push
      R(q), and stay in expand
    – If a is not the second symbol of R(q) and q is unmarked, then reject

reject:
    – Stay in reject
```

**Table 1.** A five-state, one-queue machine that recognizes $L_\infty$.

The expand and expand2 states are a little tricky. When we entered them from repeat, we have $w_n$ on the stack, whose last symbol differed from the input. We then spend the next $2^{n+1}$ input symbols expanding $w_n$ to $w_{n+1}$. If the input matches this to the end, that means that we are in a block of $w_{n+1}$'s, so we keep $w_{n+1}$ on the stack and switch back to repeat. If, on the other hand, the last symbol of the input differs from $w_{n+1}$, then the $w_{n+1}$ we saw was actually just the first half of a $w_{n+2}$, and we should stay in expand. This $w_{n+2}$ is either the first one of a block of $w_{n+2}$'s or the first half of a $w_{n+3}$, and so on.

Finally, the reject state simply remembers that something went wrong along the way. We give an example of the machine's operation in table 2.

Just as push-down automata correspond to context-free grammars, queue automata correspond to *breadth-first context-free* grammars [3]. Their production rules are of the form $v \to \alpha\beta$, where $v$ is a variable, $\alpha$ is a string of terminals and $\beta$ is a string of variables. But rather than inserting $\alpha\beta$ where $v$ appears in a derived string, we insert $\alpha$ at $v$'s position and append $\beta$ to the end of the string:

$$xvy \to x\alpha y\beta$$

If we always apply the production rule to the leftmost (least recently produced) variable, the derivation mimics the operation of a queue automaton, where the string of variables represents the queue and the growing string of terminals represents the part of the input that has been read so far. Then any language $L$ recognized by a one-queue automaton can be written

$$L = h(L_{\text{BCF}} \cap R)$$

where $L_{\text{BCF}}$ is breadth-first context-free, $R$ is regular, and $h$ is a a *homomorphism* that maps symbols to strings [4].

| input symbol | state after reading input | queue |
|---|---|---|
|   | zero | () |
| 1 | repeat | $(1')$ |
| 1 | repeat | $(1')$ |
| 0 | expand | $(1')$ |
| 1 | expand2 | $(1')$ |
| 1 | expand | $(1, 0')$ |
| 1 | expand2 | $(1, 0')$ |
| 0 | expand | $(0', 1, 0)$ |
| 1 | expand2 | $(0', 1, 0)$ |
| 1 | repeat | $(1, 0, 1, 1')$ |
| 1 | repeat | $(0, 1, 1', 1)$ |
| 0 | repeat | $(1, 1', 1, 0)$ |
| 1 | repeat | $(1', 1, 0, 1)$ |
| 0 | expand | $(1, 0, 1, 1')$ |

**Table 2.** An example of the queue automaton's operation. Here we accept the initial string $1\,1011\,1011\,1010 = w_0 w_2 w_3$.

The idea is that $L_{\text{BCF}} \cap R$ is a language in a decorated alphabet, whose variables are the queue symbols and whose terminals include both the input symbol and the machine's internal state. Then $R$ enforces the proper transitions in the finite-state control, $L_{\text{BCF}}$ keeps track of the queue, and $h$ removes the decorations and leaves just the input. If the automaton operates in real time, with one step per input symbol, then $h$ is *alphabetic* or *length-preserving*, mapping symbols to single symbols.

We can easily transform the above queue automaton into a BCF grammar. In addition to marks at the end of the $w_n$, our decorated alphabet will include subscripts $z$, $r$ and $e$ to indicate whether the machine was in the zero, repeat or expand states when the input was read, and subscripts $r \to e$ and $e \to r$ to indicate transitions. Then with our variables in bold, we have:

$$\begin{aligned}
\mathbf{I} &\to 0_z \mathbf{I},\ 1'_r \mathbf{1}' \\
\mathbf{0} &\to 0_r \mathbf{0},\ 1_e 1_e \mathbf{11} \\
\mathbf{1} &\to 1_r \mathbf{1},\ 1_e 0_e \mathbf{10} \\
\mathbf{0}' &\to 0'_r \mathbf{0}',\ 1'_{r \to e} \mathbf{0}',\ 1_e 1'_{e \to r} \mathbf{11}',\ 1_e 0'_e \mathbf{11}' \\
\mathbf{1}' &\to 1'_r \mathbf{1}',\ 0'_{r \to e} \mathbf{1}',\ 1_e 0'_{e \to r} \mathbf{10}',\ 1_e 1'_e \mathbf{10}'
\end{aligned}$$

Then let $R$ be the regular language

$$R = 0_z^* \left( (A_r' \; + \; A_{r \to e}' A_e^* (A_e' A_e^*)^* A_{e \to r}') \, A_r^* \right)^*$$

where $A_r' = \{0_r', 1_r'\}$ and so on. This ensures that transitions from $r$ to $e$ and back only occur when reading a marked symbol, and checks that the transitions are consistent. Finally, let $h$ erase all marks and subscripts.

This grammar produces growing strings of terminals with a growing queue. The reader can easily add additional subscripts indicating that no new variables will be generated, and add a condition to $R$ that makes sure the input word ends with these terminals. This corresponds to simulating our deterministic queue automaton with a non-deterministic one, which guesses when to stop pushing symbols and accepts with an empty queue.

### 3.4 Two stacks

In this section, we will show that $L_\infty$ can be recognized by a deterministic real-time automaton with access to two stacks. Equivalently, $L_\infty$ can be written

$$L_\infty = h(L_1 \cap L_2)$$

where $L_1$ and $L_2$ are context-free languages. With no time restriction, two stacks are sufficient to simulate a universal Turing machine; restricting our automaton real-time means that $h$ is again an alphabetic homomorphism.

We use the fact that the $w_n$'s are palindromes except for their last symbol. If we again mark the symbols on which the queue cycles through to the last symbol of some $w_n$, any word in $L_\infty$ can be written

$$0^* a_0' v_0 a_1' v_1 a_2' v_2 \cdots$$

where each $a_i'$ is $0'$ or $1'$, and each $v_i$ is a palindrome of 0s and 1s.

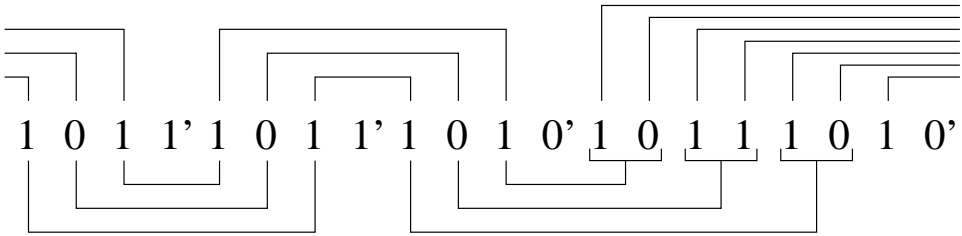

**Fig. 3.** Using two context-free languages to `repeat` and `expand` from $v_i$ to $v_{i+1}$. One language takes care of pairs where $i$ is even, the other where $i$ is odd.

The reader can check that $v_{i+1}$ is a copy of $v_i$ if $a_i' = a_{i+1}'$, and a renormalized copy $R(v_i)$ followed by 1 if $a_i' \neq a_{i+1}'$. Using the fact that $\overline{v}_i = v_i$ where $\overline{v}$ denotes

$v$ in reverse, if we define the context-free language

$$L = \left\{ a'_1 v_1 a'_2 v_2 \,\middle|\, \begin{array}{ll} v_2 = \overline{v}_1 & \text{if } a'_1 = a'_2 \\ v_2 = R(\overline{v}_1)\, 1 & \text{if } a'_1 \neq a'_2 \end{array} \right\}$$

and

$$L_1 = \mathrm{INIT}(0^* L^*)$$
$$L_2 = \mathrm{INIT}(0^* 1' L^*)$$

then $L_1$ and $L_2$ enforce this relationship between $v_i$ and $v_{i+1}$ for even and odd $i$ respectively. As an added bonus, $L_2$ makes sure $a'_0 = 1'$ and $v_0 = \epsilon$. If $h$ removes marks as before, then $L_\infty = h(L_1 \cap L_2)$.

The deterministic two-stack automaton this corresponds to is almost identical to the queue automaton of the previous section. We store the current $v_i$ on one of the stacks. To `repeat` or `expand` we pop it off one stack and push it on the other, with or without renormalization. This reverses the order of the symbols, but that doesn't matter since the $v_i$ are palindromes. We double the number of states in order to keep track of which stack we're currently reading from, and double them again to remember the last symbol $a'_i$ of the previous $w_n$; we compare this to the input $a'_{i+1}$ and switch from `repeat` to `expand` if they don't match, and so on. The reader can easily fill in the details.

## 4 Transcendentality

### 4.1 The growth function

There are several functions we can associate with a language. One is the *growth function* $N(l)$, the number of allowed words of length $l$. For regular languages, the leading behavior of $N$ is of the form $l^k \lambda^l$ where $k$ is an integer and $\lambda$ is algebraic. Unambiguous context-free languages have the same form, except that $k$ may be rational [8]. For instance, the growth function of the Dyck language is the sequence of Catalan numbers, $N(2l) = \frac{1}{l+1} \binom{2l}{l} \approx l^{-3/2} 4^l / \sqrt{\pi}$.

Throughout this section, we will refer to $\mathrm{INIT}(L_\infty)$ simply as $L_\infty$. To calculate $N(l)$ for $L_\infty$, consider two of its subsets,

$$L' = \mathrm{INIT}(1^* (10)^* (1011)^* \cdots)$$

i.e. the set of words in $L_\infty$ that don't start with 0, and

$$L'' = \mathrm{INIT}((10)^* (1011)^* \cdots)$$

i.e. the set of words in $L_\infty$ that don't start with 0 or 11. Let $N'(l)$ and $N''(l)$ be the growth functions of $L'$ and $L''$ respectively.

Then note that any word in $L_\infty$ consists of a word in $L'$ preceded by none or more 0s, and similarly any word in $L'$ consists of a word in $L''$ preceded by none or more 1s. That is, $L_\infty = 0^*L'$ and $L' = 1^*L''$. Then

$$N(l) = \sum_{j=0}^{l} N'(j)$$

$$N'(l) = \sum_{j=0}^{l} N''(j) - 1 \quad \text{for } l > 0 \tag{1}$$

(The $-1$ here prevents double-counting of the word $1^l$, since both 1 and $\epsilon$ are members of $L''$. Obviously, we can ignore it for large $l$.) In addition, the renormalization $R$ gives a one-to-one correspondence between words in $L'$ and words in $L''$ that are twice as long, so

$$N'(l) = N''(2l) \tag{2}$$

Now if we approximate (1) as an integral, combine it with (2), and differentiate both sides, we get

$$dN'(l)/dl = N'(l/2)$$

Solving this with series yields

$$N'(l) \propto \sum_{n=0}^{\infty} \frac{l^n}{n!\, 2^{n(n-1)/2}}$$

which clearly converges. Its leading behavior for large $l$ is

$$N'(l) \propto e^{A(\ln l)^2 + B \ln l\, \ln \ln l + C \ln l} = l^{A \ln l}\, l^{B \ln \ln l}\, l^C$$

where

$$A = \frac{1}{2 \ln 2}$$
$$B = -\frac{1}{\ln 2}$$
$$C = \frac{1}{2} + \frac{1}{\ln 2} + \frac{\ln \ln 2}{\ln 2}$$

To obtain $N(l)$ we just integrate $N'(l)$, which changes $C$ to $C + 1$.

This leading behavior of $l^{A \ln l}$ grows faster than any power law, but slower than an exponential.

### 4.2 The generating function

Another function we can associate with a language is its *generating function*

$$G(z) = \sum_{l=0}^{\infty} N(l) \, z^l$$

$G(z)$ is rational for regular languages, and algebraic for unambiguous context-free languages [5].

As before, consider $L'$ and $L''$, and let their generating functions be $G'$ and $G''$ respectively. Equation (1) now becomes

$$G(z) = (1 + z + z^2 + \cdots) \, G'(z) = \frac{G'(z)}{1 - z}$$

$$G'(z) = (1 + z + z^2 + \cdots) \, G''(z) - (z + z^2 + \cdots) = \frac{G''(z) - z}{1 - z} \quad (3)$$

In addition, every word in $L''$ of even length is a renormalized word in $L'$, and every word in $L''$ of odd length is a renormalized word in $L'$ followed by 1. This gives

$$G''(z) = (1 + z) \, G'(z^2)$$

which when combined with (3) gives

$$G'(z) = \frac{1 + z}{1 - z} G'(z^2) - \frac{z}{1 - z}$$

This can be simplified somewhat by writing $G'(z) = \frac{g(z)}{1-z} + 1$. Then

$$g(z) = z + \frac{1}{1 - z} g(z^2) \quad (4)$$

Using the fact that $g(0) = 0$ since $G'(0) = 1$, this gives

$$g(z) = z + \frac{1}{1 - z}(z^2 + \frac{1}{1 - z^2}(z^4 + \frac{1}{1 - z^4}(\cdots)))$$

$$= z + \frac{1}{1 - z} z^2 + \frac{1}{1 - z} \frac{1}{1 - z^2} z^4 + \cdots$$

Since $N(l)$ grows less than exponentially, $G(z)$ will converge for $|z| < 1$ and diverge at $|z| = 1$. We're particularly interested in the nature of that divergence. If $z = 1 - \epsilon$ where $\epsilon$ is small, (4) becomes

$$f(\epsilon) \approx \epsilon^{-1} f(2\epsilon)$$

where $f(\epsilon) = g(1 - \epsilon)$. As $\epsilon$ goes to zero, this gives a leading behavior of

$$f(\epsilon) \propto \epsilon^{A \ln \epsilon} \epsilon^{-B}$$

with $A = 1/(2 \ln 2)$ and $B = 1/2$. Then $G'(1 - \epsilon)$ and $G(1 - \epsilon)$ diverge with the same form, except that $B = 3/2$ and $B = 5/2$ respectively.

There are two ways to see that $G$ is transcendental. The first is that the poles of an algebraic function must diverge polynomially, and $\epsilon^{A\ln\epsilon}$ diverges faster than any polynomial. The other is that there is a pole at $z = e^{i\theta}$ whenever $\theta$ is a rational fraction of $2\pi$, since these points are periodic under the transformation $z \to z^2$. Thus the unit circle has a dense set of poles and forms an essential boundary, outside which $G$ cannot be analytically continued.

### 4.3 The dynamical zeta function

Many time-averaged properties of dynamical systems, such as Lyapunov exponents, repeller lifetimes, etc. can be calculated from a generating function for the set of periodic orbits of the system. We can write this as a sum over the set of prime cycles $\{p\}$; since eacn one can start at $p$ different points, and can be repeated any number of times, we have

$$G(z) = \sum_{\{p\}} = \frac{p\,z^p}{1 - z^p}$$

This can be written as a logarithmic derivative of a *dynamical zeta function* $\zeta$ [7, 17]:

$$G(z) = z\frac{d}{dz}\log\zeta$$

where

$$\zeta^{-1}(z) = \prod_{\{p\}}(1 - z^p)$$

For the one-humped map at the transition to chaos, we have one prime cycle of period $2^k$ for each $k$, and another of period 1, so

$$\zeta^{-1}(z) = (1 - z)\,g(z)$$

where

$$g(z) = (1 - z)(1 - z^2)(1 - z^4)\cdots = (1 - z)\,g(z^2)$$

giving $\zeta$ a pole at $e^{i\theta}$ whenever $\theta$ is a dyadic rational fraction of $2\pi$. Letting $g(1 - \epsilon) = f(\epsilon)$ gives a similar recurrence as before,

$$f(\epsilon) \approx \epsilon f(2\epsilon) \tag{5}$$

giving

$$\zeta(1 - \epsilon) \propto \epsilon^{A\ln\epsilon}\epsilon^B$$

with, once again, $A = 1/(2\ln 2)$ and $B = -3/2$. This divergence for the zeta function was pointed out in [13].

## 5  Conclusion

We have studied the period-doubling transition to chaos from a computational and analytical point of view. Since one queue suffices to recognize its symbolic dynamics in real time, but two stacks are necessary, we argue that the system's long-term memory has a first-in, first-out character. We have also shown that its growth, generating, and dynamical zeta functions have an interesting transcendental form.

To what extent are these things common to other dynamical phase transitions? For period-multiplying cascades, the transcendental forms remain the same. If we have a series of orbits of period $r^k$ for some $r$ other than 2, the recurrence (5) becomes

$$f(\epsilon) = \epsilon f(r\epsilon)$$

giving exactly the same divergence for $\zeta$, except that now $A = 1/(2\ln r)$. The growth and generating functions behave similarly.

As far as grammatical structure is concerned, the indexed context-free grammar and queue automaton given above can clearly be adapted to a wide variety of renormalizations. But how general is our construction of a two-stack automaton? The Morse sequence $0110\,1001\,1001\,0110\cdots$ generated by $0 \to 01$, $1 \to 10$ is its complement in reverse, so a two-stack machine can easily be constructed for it. On the other hand, the Fibonacci sequence $abaab\,aba\,abaab\cdots$ generated by $a \to ab$, $b \to a$, which occurs in the quasiperiodic behavior of coupled oscillators, can be recognized with a queue [4] but does not seem to have any obvious palindromic or stack-like structure. We leave this as an open question.

**Acknowledgements.** The section on transcendentality was done some time ago during visits at the Nonlinear Systems Laboratory of the University of Warwick, NORDITA in Copenhagen, and the Institute for Theoretical Physics in Göteborg. C.M. thanks these institutions for their hospitality; Predrag Cvitanović, Mark Newman, Marek Chrobak, Eric Aurell and Kunihiko Kaneko for helpful conversations; Molly Rose and Spootie the Cat for their companionship; and the steamboat Bohuslän for the proof of lemmas 1-6 in the Appendix. This work was supported in part by NSF grant ASC-9503162.

## A  Appendix: $L_\infty$ is not context-free

In this Appendix, show that $L_\infty$ and $\text{INIT}(L_\infty)$ are not context-free, and prove a number of enjoyable combinatorial results about these languages on the way.

We will use another useful characterization of the $w_n$,

$$w_0 = 1$$
$$w_1 = 10$$
$$w_n = w_{n-1} w_{n-2}^2 \text{ for } n \geq 2$$

Iterating this gives us a useful equation:

$$w_n = w_i w_{i-1}^2 w_i^2 w_{i+1}^2 \cdots w_{n-2}^2 \text{ for } 0 < i < n \qquad (6)$$

Then we can define $L_\infty$ as

$$L_\infty = \{w_{i_1} w_{i_2} w_{i_3} \cdots w_{i_m} : i_1 \leq i_2 \leq \cdots \leq i_m,\ 0 \leq m\}$$

We will call initial and final substrings *heads* and *tails* respectively, and call $\beta$ a *proper head* of $\alpha$ if $\beta = \alpha\gamma$ where $\gamma$ is non-empty, i.e. $|\beta| < |\alpha|$. Then to describe words in $\text{INIT}(L_\infty)$, we use the following lemma:

**Lemma 1.** *If $\beta$ is a proper head of $w_n$, then it can be decomposed into $w_i$ of strictly decreasing size, i.e. $\beta = w_{i_1} w_{i_2} \cdots w_{i_m}$ where $n > i_1 > i_2 > \cdots > i_m$ for some $m \geq 0$. Therefore, $\text{INIT}(L_\infty)$ can be written*

$$\text{INIT}(L_\infty) = \{w_{i_1} w_{i_2} w_{i_3} \cdots w_{i_m} :$$
$$i_1 \leq i_2 \leq \cdots \leq i_k > i_{k+1} > \cdots > i_m, \ 0 \leq k \leq m\} \qquad (7)$$

*with decompositions consisting of a non-decreasing part followed by a strictly decreasing part.*

*Proof.* The proof is by induction on $n$. The first half of $w_n$ is $w_{n-1}$, and the second half of $w_n$ only differs from $w_{n-1}$ on its last symbol. Therefore, if $\beta$ is a proper head of $w_n$, we have two cases: if $|\beta| < 2^{n-1}$, then $\beta$ is a proper head of $w_{n-1}$, and if $|\beta| \geq 2^{n-1}$, then $\beta = w_{n-1} \beta'$ where $\beta'$ is a proper head of $w_{n-1}$. Since the only proper head of $w_0$ is the empty string, the induction is complete.

To get equation (7), note that any word in $\text{INIT}(L_\infty)$ can be written $w_{i_1} \cdots w_{i_k} \beta$ where $i_1 \leq \cdots \leq i_k$ and $\beta$ is a (possibly empty) proper head of $w_{i_{k+1}}$. □

We will also find the following useful:

**Lemma 2.** *If $\beta$ is a proper tail of $w_n$, then it can be decomposed into $w_i$ of non-decreasing size, i.e. $\beta = w_{i_1} w_{i_2} \cdots w_{i_m}$ where $i_1 \leq i_2 \leq \cdots \leq i_m < n$ for some $m \geq 0$.*

*Proof.* We use a induction similar to that in the previous lemma. Let $\beta$ be a proper tail of $w_n$. Recall that $w_n = w_{n-1} w_{n-2}^2$. If $|\beta| < 2^{n-2}$, then $\beta$ is a proper tail of $w_{n-2}$. If $2^{n-2} \leq |\beta| < 2^{n-1}$, then $\beta = \beta' w_{n-2}$ where $\beta'$ is a proper tail of $w_{n-2}$. Finally, if $2^{n-1} \leq |\beta| < 2^n$, then $\beta = \beta' w_{n-2}^2$ where $\beta'$ is a proper tail of $w_{n-1}$.

Thus every tail has a non-decreasing decomposition, in which, moreover, each $w_i$ is repeated at most twice. □

We now prove an important lemma. Note that if $\alpha$ is in $\text{INIT}(L_\infty)$, so are all its substrings. Call a head or tail a *w-head* or *w-tail* if it consists of some $w_i$.

**Lemma 3.** *The decomposition of words in $\text{INIT}(L_\infty)$ given in equation (7) is unique.*

*Proof.* Call a decomposition $\alpha = w_{i_1} w_{i_2} w_{i_3} \cdots w_{i_m}$ *legal* if it is first non-decreasing and then strictly decreasing, $i_1 \leq \cdots \leq i_k > i_{k+1} > \cdots > i_m$ for some $k \leq m$. We wish to show that no word has more than one legal decomposition. To prove this by contradiction, let $\alpha$ be a word of minimal length with two distinct legal decompositions,

$$\alpha = w_{i_1} w_{i_2} \cdots w_{i_m}$$
$$= w_{j_1} w_{j_2} \cdots w_{j_n}$$

Since $\alpha$ has minimal length, these must differ on the first block, i.e. $i_1 \neq j_1$. Otherwise, $\alpha = w_{i_1} \beta$ where $\beta$ is a shorter word with two legal decompositions.

We assume without loss of generality that $i_1 < j_1$ and apply equation (6) to the second decomposition, obtaining

$$\alpha = w_{i_1} w_{i_2} \cdots w_{i_m}$$
$$= w_{i_1} w_{i_1-1}^2 w_{i_1}^2 \cdots w_{j_1-2}^2 w_{j_2} \cdots w_{j_n}$$

Then if $\alpha = w_{i_1}\beta$, we have

$$\beta = w_{i_2} \cdots w_{i_m}$$
$$= w_{i_1-1}^2 w_{i_1}^2 \cdots w_{j_1-2}^2 w_{j_2} \cdots w_{j_n}$$

Both these decompositions are legal. If $i_1 \leq i_2$, we have $i_1 - 1 < i_2$, and they differ on the first block. If $i_2 < i_1$ then either $i_2 < i_1 - 1$ and they differ on the first block, or $i_2 = i_1 - 1$. But in this case $\alpha$'s first decomposition is decreasing all the way, so $i_3 < i_2$ and they differ on the second block. In every case, $\beta$ is a shorter word with two legal decompositions, again contradicting our assumption that $\alpha$ has minimal length. The proof is completed by noting that the decomposition of the empty word is unique! $\square$

This uniqueness has the following corollary.

**Lemma 4.** *Each of the $w_i$ in $\alpha$'s legal decomposition is the largest $w$-head of the tail of $\alpha$ starting with it, and the largest $w$-tail of the head of $\alpha$ ending with it. Furthermore, any $w_n$ that occurs as a subword in $\alpha$ must be contained in some $w_i$ in $\alpha$'s decomposition; it cannot cross the boundaries between the $w_i$.*

*Proof.* To show that $w_{i_1}$ is the largest $w$-head of $\alpha$, suppose that $w_n$ is a $w$-head of $\alpha$ where $n > i_1$. Then $w_n = w_{i_1} \cdots w_{i_l}\beta$ for some $l$, where $\beta$ is a head of $w_{i_{l+1}}$. Then using lemma 1 on $\beta$ gives a legal decomposition for $w_n$ other than $w_n$ itself (note that the decomposition given by equation (6) is not legal). Therefore, any $w$-head of $\alpha$ must be contained in $w_{i_1}$. By induction, each of the $w_i$ is the largest $w$-head of the part of $\alpha$ starting with it.

On the other end, we wish to show that $w_{i_m}$ is the largest $w$-tail of $\alpha$. If $w_n$ is a $w$-tail of $\alpha$ where $n > i_m$, then $w_n = \beta w_{i_l} \cdots w_{i_m}$ for some $l$, where $\beta$ is a tail of $w_{i_{l-1}}$. Using lemma 2 on $\beta$ gives a legal decomposition for $w_n$ other than itself, and we again have a contradiction. By induction, each $w_i$ is the largest $w$-tail of the part of $\alpha$ ending with it.

Finally, suppose some subword $w_n$ of $\alpha$ is not contained in any of the $w_i$'s of $\alpha$'s decomposition, and that it crosses one or more boundaries between the $w_i$. Then $w_n = \beta w_{i_l} \cdots w_{i_p}\gamma$ where $\beta$ is a tail of $w_{i_{l-1}}$ and $\gamma$ is a head of $w_{i_{p+1}}$. Using lemmas 1 and 2 again gives a decomposition for $w_n$ other than itself. $\square$

This means that given $\alpha \in \text{INIT}(L_\infty)$, we can find $\alpha$'s legal decomposition by proceeding from the left (or the right) and taking the largest $w$-head (or $w$-tail) at each step. Note that the uniqueness of this decomposition is also implicit in our calculation of the generating function for $\text{INIT}(L_\infty)$ in section 4.2.

Next, we prove that the $w_n$ don't have square heads, cubic tails, or quartic subwords.

**Lemma 5.** *No $w_n$ has a head of the form $\alpha^2$, a tail of the form $\alpha\beta\alpha\beta\alpha$, or a subword of the form $\alpha^4$, where $|\alpha| > 0$.*

*Proof.* We prove the first statement first. It is true for $w_0$. Suppose $w_n = \alpha\alpha\delta$, and assume that no smaller $w_n$ can be written this way. We must have $|\alpha| > 2^{n-2}$ since otherwise $w_{n-1}$ has $\alpha\alpha$ for a head. Therefore, the middle of $w_n$ falls inside the second $\alpha$, dividing it into two subwords $\alpha = \beta\gamma$. Then $w_n = \beta\gamma\beta\gamma\delta$, and since $w_n$'s first and second halves are identical except for the last symbol, we have $w_{n-1} = \beta\gamma\beta = \gamma\tilde{\delta}$, where $\tilde{v}$ denotes a word $v$ with the last symbol complemented.

We now have two cases. If $|\gamma| \geq |\beta|$, write $\gamma = \beta\zeta$ and $w_{n-1} = \beta\beta\zeta\beta$ has a square head. If $|\beta| > |\gamma|$, write $\beta = \gamma\zeta$ and $w_n = \gamma\zeta\gamma\gamma\zeta$. The middle of $w_n$ must divide the second $\gamma$ into $\gamma = \eta\theta$, so $w_{n-2} = \eta\theta\zeta\eta = \theta\eta\theta\tilde{\zeta}$. If $|\theta| \geq |\eta|$, write $\theta = \eta\kappa$ and $w_{n-2} = \eta\eta\kappa\zeta\eta$; if $|\eta| > |\theta|$, write $\eta = \theta\kappa$ and $w_{n-2} = \theta\theta\kappa\theta\tilde{\zeta}$. In either case, $w_{n-2}$ has a square head. By contradiction, no $w_n$ can have a square head.

To show that no $w_n$ can have a tail of the form $\alpha\beta\alpha\beta\alpha$ where $|\alpha| > 0$, recall that $w_n$ with its last symbol removed is a palindrome. Suppose that $w_n = \delta\alpha\beta\alpha\beta\alpha$ and $\alpha = \gamma a$ where $a$ is $\alpha$'s last symbol. Then $w_n = \delta\gamma a\beta\gamma a\beta\gamma a = \overline{\gamma}\overline{\beta}a\overline{\gamma}\overline{\beta}a\overline{\gamma}\overline{\delta}a$, and $w_n$ has a square head $(\overline{\gamma}\overline{\beta}a)^2$ where $\overline{v}$ denotes a word $v$ written in reverse.

Finally, suppose $w_n$ has a quartic subword, $w_n = \beta\alpha^4\gamma$, and no smaller word does. Then the middle of $w_n$ must divide one of the $\alpha$'s into $\alpha = \zeta\eta$, where without loss of generality $\zeta$ is non-empty. Then $w_n = \beta\zeta \cdot \eta\zeta \cdot \eta\zeta \cdot \eta\zeta \cdot \eta\gamma$ where the middle of $w_n$ falls on one of the four $\cdot$'s. The first two give $w_{n-1}$ a square head $\eta\zeta\eta\zeta$, and the last two give $w_{n-1}$ a tail $\zeta\eta\zeta\eta\zeta$. By contradiction, no $w_n$ can have a subword of the form $\alpha^4$. □

We now prove a result which will be instrumental in showing that words in INIT($L_\infty$) can only be pumped in a limited way. Call $\alpha$ a *w-cycle* if it is a cyclic permutation of a non-zero power of some $w_n$, i.e. $\alpha = \epsilon$ or $\beta w_n^p \gamma$ for some $p \geq 0$ where $\gamma\beta = w_n$.

**Lemma 6.** *If $\alpha$, $\alpha^2$ and $\alpha^3$ are in INIT($L_\infty$), then $\alpha$ is a w-cycle.*

*Proof.* Let $\alpha$ and $\alpha^2$ have legal decompositions

$$\alpha = w_{i_1} w_{i_2} \cdots w_{i_k} \cdots w_{i_m}$$
$$\alpha^2 = w_{i_1} w_{i_2} \cdots w_{i_k} \cdots w_{i_m} \cdot w_{i_1} w_{i_2} \cdots w_{i_k} \cdots w_{i_m}$$
$$= w_{j_1} w_{j_2} \cdots w_{j_l} \cdots w_{j_n}$$

where $i_1 \leq i_2 \leq \cdots \leq i_k > \cdots > i_m$ and $j_1 \leq j_2 \leq \cdots \leq j_l > \cdots > j_n$.

First, note that $i_p = j_p$ for $p < k$, so that $w_{i_1} \cdots w_{i_{k-1}}$ starts the legal decomposition of $\alpha^2$. If not, the first $w_{j_p}$ for which this is not true, or the head of it that overlaps the first $\alpha$, has a decomposition starting $w_{i_p} w_{i_{p+1}} \cdots$ where $i_p \leq i_{p+1}$. This would contradict lemma 1, which shows that the legal decomposition of a head of $w_{j_p}$ is strictly decreasing.

Similarly, $i_p = j_{p-m+n}$ for $p > k$, so that $w_{i_{k+1}} \cdots w_{i_m}$ concludes the legal decomposition of $\alpha^2$. Otherwise, the last $w_{j_p}$ for which this is not true, or its tail that overlaps the second $\alpha$, has a decomposition ending $\cdots w_{i_{p-1}} w_{i_p}$ where $i_{p-1} > i_p$, whereas by lemma 2 its decomposition must be non-decreasing.

Therefore, we write
$$\alpha^2 = w_{i_1} \cdots w_{i_{k-1}} \, \beta \, w_{i_{k+1}} \cdots w_{i_m}$$
where
$$\beta = w_{i_k} \cdots w_{i_m} \cdot w_{i_1} \cdots w_{i_k}$$
$$= w_{j_k} \cdots w_{j_{k-m+n}}$$
where $w_{j_k} \cdots w_{j_{k-m+n}}$ is $\beta$'s legal decomposition. We will now explore a series of cases regarding the structure of $\beta$.

**Case I:** $j_k = i_k$ and $j_{k-m+n} \neq i_k$. We must have $j_{k+1} \geq j_k = i_k$. Otherwise, $\beta$'s decomposition starts out decreasing, and its length is a sum of decreasing powers of 2 totalling less than $2 \cdot 2^{i_k}$, which contradicts the fact that it starts and ends with $w_{i_k}$. Since $i_{k+1} < i_k$, we have $j_{k+1} > i_{k+1}$, and so $w_{j_{k+1}}$ must overlap the second $\alpha$ to avoid having a legal (decreasing) decomposition other than itself. Similarly, $j_{k-m+n} > i_k$ by lemma 4, and $w_{j_{k-m+n}}$ must overlap with the first $\alpha$ to avoid having a non-decreasing decomposition.

Since this means $w_{j_{k+1}}$ and $w_{j_{k-m+n}}$ overlap, they must coincide. Then $\beta = w_{i_k} w_{j_{k+1}}$ and $\alpha$ is a cyclic permutation of $w_{j_{k+1}} = w_{i_{k+1}} \cdots w_{i_m} \cdot w_1 \cdots w_{i_k}$.

**Case II:** $j_k \neq i_k$ and $j_{k-m+n} \neq i_k$. By lemma 4 we have $j_k > i_k$, so $w_{j_k}$ must overlap with the second $\alpha$ to avoid having a decreasing decomposition. As in Case I, $w_{j_{k-m+n}}$ must overlap with the first $\alpha$ to avoid having a non-decreasing decomposition. Therefore $w_{j_k}$ and $w_{j_{k-m+n}}$ coincide, and $\beta = w_{j_k}$.

However, this leads to a contradiction when we consider $\alpha^3$. (For instance, if $\alpha = 110101$, then $\beta = 10111010 = w^3$, and $\alpha$ and $\alpha^2$ are in $\text{INIT}(L_\infty)$, but $\alpha^3$ is not.) By the same argument as for $\alpha^2$, the legal decomposition of $\alpha^3$ begins with $w_{i_1} \cdots w_{i_{k-1}}$ and ends with $w_{i_{k+1}} \cdots w_{i_m}$, so we can write
$$\alpha^3 = w_{i_1} \cdots w_{i_{k-1}} \, \beta \, \delta \, w_{i_{k+1}} \cdots w_{i_m}$$
where
$$\delta = w_{i_{k+1}} \cdots w_{i_m} \cdot w_{i_1} \cdots w_{i_k}$$
and $\beta = w_{i_k} \delta$. Since $\delta$ is a tail of $\beta = w_{j_k}$, by lemma 2 its decomposition is non-decreasing. On the other hand, the legal decomposition of $\alpha^3$ must include $\beta = w_{j_k}$ and then decrease, since otherwise the block starting with $\beta$, or the next one, would run past the end of $\delta$. Since $\delta$'s decomposition is both decreasing and non-decreasing, it must be a single $w_s$ for some $s$; but this gives $\beta = w_{j_k} = w_{i_k} w_s$, a contradiction.

**Case III:** $j_k \neq i_k$ and $j_{k-m+n} = i_k$. As in Case II, $j_k > i_k$ and $w_{j_k}$ must overlap the second $\alpha$. Let $q$ be the rightmost index such that $j_q \neq i_{q+m-n}$. Then $j_q > i_{q+m-n}$ by lemma 4, so $w_{j_q}$ must overlap the first $\alpha$, and $w_{j_k}$ and

$w_{j_q}$ coincide. Moreover, the sequence $j_{q+1} = i_{q+m-n+1}, \ldots, j_{k-m+n} = i_k$ is non-decreasing, but it must also be decreasing since $j_k > i_k$. Thus $q = k - m + n - 1$ and this sequence consists only of $j_{k-m+n} = i_k$, so $\beta = w_{j_k} w_{i_k}$ and $\alpha$ is a cyclic permutation of $w_{j_k} = w_{i_k} \cdots w_m \cdot w_1 \cdots w_{i_{k-1}}$.

**Case IV:** $j_k = i_k$ and $j_{k-m+n} = i_k$. As in Case I, $j_{k+1} > i_{k+1}$ and $w_{j_{k+1}}$ must overlap the second $\alpha$. As in the previous case, let $q$ be the rightmost index such that $j_q \neq i_{q+m-n}$. Then $w_{j_q}$ must overlap the first $\alpha$, so $w_{j_{k+1}}$ and $w_{j_q}$ coincide. Moreover, the sequence $j_{q+1} = i_{q+m-n+1}, \ldots, j_{k-m+n} = i_k$ is non-decreasing, so these must all be equal to $i_k$ since $j_k = i_k$. Thus $\beta = w_{i_k} w_{j_q} w_{i_k}^p$ where $p = k - m + n - q \geq 1$. If $p > 1$, we must have $j_q = i_k$ for this decomposition to be legal, $\beta = w_{i_k}^{p+2}$, and $\alpha$ is a cyclic permutation of $w_{i_k}^{p+1}$.

If $p = 1$, we need to examine $\alpha^3$ to eliminate the possibility that $\beta = w_{i_k} w_{j_q} w_{i_k}$ where $j_q > i_k$. (For instance, if $\alpha = 011101$, then $\beta = 10\,1011\,10 = w_1 w_2 w_1$, $\alpha$ and $\alpha^2$ are in $\text{INIT}(L_\infty)$, but $\alpha^3$ is not.) This would give us

$$\alpha^3 = w_{i_1} \cdots w_{i_{k-1}} \gamma \, w_{i_{k+1}} \cdots w_{i_m}$$

where $\gamma = w_{i_k} w_{j_q} w_{i_k} w_{j_q} w_{i_k}$, or $\gamma = w_i w_j w_i w_j w_i$ for short. We will show that no such word with $j > i$ is in $\text{INIT}(L_\infty)$.

Suppose the largest $w$-head of $\gamma$ is some $w$ larger than $w_i$. By lemma 4 the $w_i$ and $w_j$ cannot be cut by $w$'s right boundary, so $w_k$ must be $w_i w_j$, $w_i w_j w_i$, $w_i w_j w_i w_j$, or $w_i w_j w_i w_j w_i$. The first two of these give $w$ a legal decomposition, while the last two give it a square head and violate lemma 5. So $w = w_i$ and $\gamma$'s legal decomposition starts with $w_i$.

Next, suppose the largest $w$-head of $w_j w_i w_j w_i$ is some $w$ larger than $w_j$. The same arguments eliminate all possibilities except $w = w_j w_i w_j$; but if $j > i$, the length of this cannot be a power of 2. So $\gamma$'s legal decomposition starts with $w_i w_j$. Since the remainder of $\gamma$ is already legally decomposed into $w_i w_j w_i$, its entire decomposition must be $w_i w_j w_i w_j w_i$; but for this to be legal, we must have $i = j$.

Thus if $p = 1$, then $j_q = i_k$, $\beta = w_{i_k}^3$ and $\alpha$ is a cyclic permutation of $w_{i_k}^2$. This completes the proof for all cases. □

We are now in a position (finally!) to prove the following:

**Theorem 7.** *Neither $L_\infty$ nor $\text{INIT}(L_\infty)$ are context-free.*

*Proof.* Ogden's Lemma [15] states that for any context-free language $L$, there is a constant $k$ such that if $k$ or more symbols in a word $\xi \in L$ are designated as 'distinguished,' then $\xi = \alpha \beta \gamma \delta \mu$ where:

- $\gamma$ contains at least one distinguished symbol
- Either $\beta$ or $\delta$ contains at least one distinguished symbol
- If $\beta$ (resp. $\delta$) contains distinguished symbols, then $\alpha$ (resp. $\mu$) does also
- $\beta \gamma \delta$ contains at most $k$ distinguished symbols
- $\alpha \beta^p \gamma \delta^p \mu \in L$ for all $p \geq 0$

We will refer to $\alpha\beta^p\gamma\delta^p\mu$ as $\xi_p$.

Suppose $\text{INIT}(L_\infty)$ is context-free. Then let $\xi$ be $w_n^5$ with the last symbol removed, i.e. $\xi = w_n^4 w_{n-1} w_{n-2} \cdots w_0$, with $n$ chosen such that $2^n > k$. Mark the last $k$ symbols of $\xi$ as distinguished; these are all to the right of the last $w_n$.

Since $\gamma$ contains a distinguished symbol, $\delta$ either consists entirely of distinguished symbols or is empty. In the first case, since $\delta$ can be pumped, it must be a $w$-cycle by lemma 6, and since $|\delta| < k < 2^n$, it must be a cyclic permutation of some $w_i$ with $i < n$. For $p \geq 5$, then, $\xi_p$ contains a quartic subword $w_i^4$. By lemma 5 this must appear in $\xi_p$'s legal decomposition, which must therefore be non-decreasing at this point. But this is a contradiction, since with 4 copies of $w_n$ to the left of $\delta$, either $\beta$ contains $w_n$ or both $\alpha$ and $\gamma$ do, and in either case $w_n$ occurs more than once in $\xi_p$ to the left of $w_i^4$.

If $\delta$ is empty, then both $\alpha$ and $\beta$ must contain distinguished symbols, in which case $\beta$ consists entirely of them. Then $\beta$ must be a cyclic permutation of some $w_i$ with $i < n$, and the same argument goes through. In both cases, $\xi_p \notin L$ for $p \geq 5$, and $\text{INIT}(L_\infty)$ is not context-free.

The context-free languages are closed under INIT [11], so since $\text{INIT}(L_\infty)$ is not context-free, $L_\infty$ cannot be either. □

We are confident that the same techniques can be applied to show that $L_\infty$ and $\text{INIT}(L_\infty)$ are not recognizable by one-way stack automata. The difficulty is to put an upper bound on the size of the $\beta_j$ and $\psi_j$ (in the notation of [16]) as we did here for $\delta$ and $\beta$.